\documentclass[final,twocolumn]{article}

\usepackage[T1]{fontenc}
\usepackage{array}
\usepackage{lmodern}
\usepackage{caption}
\usepackage{subcaption}
\usepackage{authblk}
\usepackage{amsmath}
\usepackage{amssymb}
\usepackage[pdftex]{graphicx}
\usepackage{multirow}
\usepackage{color}
\usepackage{hyperref}
\DeclareMathOperator\erf{erf}

\title{Functional Optimization for a Beam Driven Plasma Neutralizer in DEMO Neutral Beam Injector}

\author[1,2]{Fabio Veronese}
\author[1,3]{Piero Agostinetti}
\author[4]{Christian Hopf}
\author[4]{Giuseppe Starnella}

\affil[1]{Consorzio RFX (CNR, ENEA, INFN, Università di Padova, Acciaierie Venete S.p.A.), Corso Stati Uniti 4, Padova, 35127, Italy}
\affil[2]{Centro Ricerche Fusione (CRF), Università di Padova, Corso Stati Uniti, 4, Padova, 35127, Italy}
\affil[3]{Institute for Plasma Science and Technology - Section of Padova, Corso Stati Uniti 4, Padova, 35127, Italy}
\affil[4]{Max-Planck-Institut für Plasmaphysik, Boltzmannstr. 2, Garching bei München, 85748, Germany}
\date{\url{https://doi.org/10.1016/j.fusengdes.2023.113663}}
\begin{document}
\twocolumn[
	\begin{@twocolumnfalse}
		\maketitle
		\begin{abstract}
		The Beam Driven Plasma Neutralizer (BDPN) has been proposed as a more efficient alternative to the gas neutralizer for negative-ion based Neutral Beam Injection (NNBI). In this paper we model the performance of an entire NNBI beamline with a BDPN. We simultaneously consider all the relevant physics and engineering aspects, the most important being the plasma density and degree of ionization inside the BDPN as a function of its geometry and feed gas flow, the geometrical transmission of the beamline, the dependence of the neutral gas distribution in the beamline on the geometry of the beamline components and gas flows, and the species evolution of the extracted D$^-$ beam through this neutral and charged particle distribution. Furthermore, we calculate the heat loads expected on the BDPN parts and on the NBI components located downstream of it and study the effect of the magnetic cusp field across the BDPN entrance on beamline transmission. While our results constitute an optimization only under the applied boundary conditions, we find that the beamline with a BDPN increases the system's wall plug efficiency by about 13\;\% to 0.34 from the 0.30 estimated for a gas neutralizer.\\
		\end{abstract}
	\end{@twocolumnfalse}
	]
\let\thefootnote\relax\footnotetext{E-mail: \url{fabio.veronese@igi.cnr.it}}
\section{Introduction}
\label{intro}

\begin{figure}
	\centering
	\includegraphics[width=0.9\columnwidth]{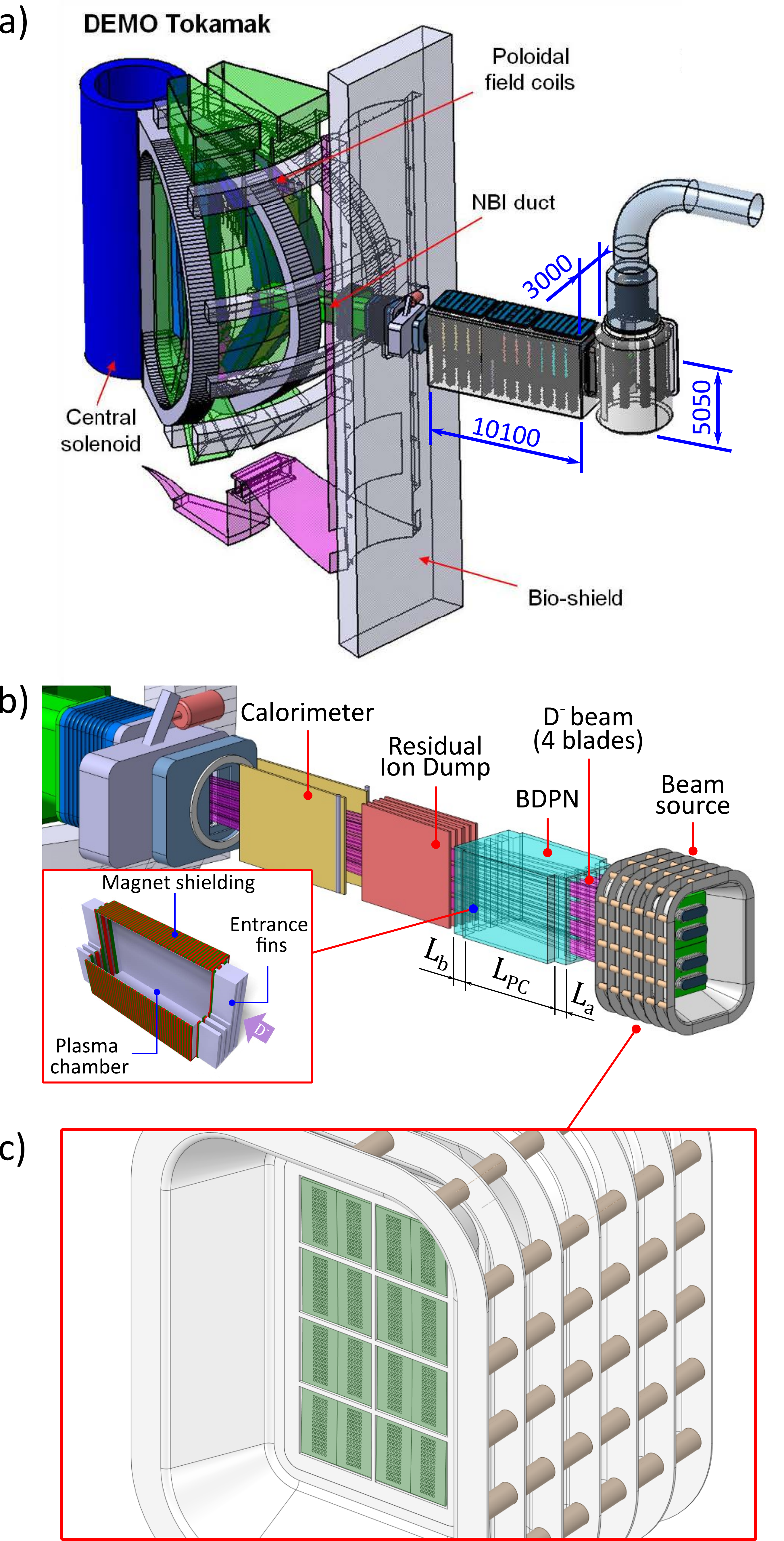}
	\caption{Pre-conceptual design for DEMO NBI (2022). a) isometric cutaway of the DEMO tokamak and NBI assembly with the main dimensions of the injector vacuum vessel; b) view of the main DEMO NBI components in the version featuring the BDPN, with a section view showing the BDPN geometry and a possible magnetic configuration \cite{starnella1}; c) detail of the beam source, showing the chosen layout of the beamlet array.}
	\label{fig:DEMO_NBI}
\end{figure}

Following in the future footsteps of ITER, a DEMOnstration fusion plant is foreseen to be the next step towards baseload power generation by thermonuclear fusion, with the aim of proving the technological and commercial feasibility of a complete power plant. To achieve this, a higher duty cycle with longer plasma discharges is required and recirculating power must be reduced by increasing efficiency of the auxiliary heating systems. For the European DEMO the pre-conceptual design \cite{mqtran1} developed in the period 2014--2020 assumes 50~MW from each candidate heating system, Electron Cyclotron Resonance Heating (ECRH), Ion Cyclotron Resonance Heating (ICRH) and Neutral Beam Injection (NBI) heating. Since 2021 the baseline design assumes ECH only, but NBI as well as ICH continue to be developed as a risk mitigation strategy. The NBI design assumes 50~MW of power injected from two or three 1~MeV beamlines with an accelerated negative deuterium ion current of 40~A.\\
The energy efficiency of NBI is limited mostly by the upper bound on neutralization for negatively charged beams through neutral gas, which forces the system to lose $\gtrsim 45\;\%$ of its power in the form of non-neutralized ions impinging on the residual ion dump. Compounded by early stripping losses, re-ionization losses, geometrical transmission losses and limited power supply efficiency these losses limit the wall-plug efficiency, i.e. injected power divided by total system power consumption, of typical negative-ion-based NBIs to about 30\;\% \cite{sonato1}.\\
A possible remedy is provided by the plasma neutralizer. In this concept the gas inside the neutralizer has a degree of ionization of several percent, enhancing the neutralizer efficiency for the negative beam \cite{berkner1}. Experiments have proven increased neutralization \cite{hanada1}, but until now they have been using external arc sources to generate plasma, introducing additional complexity and power consumption.\\
In 2013 E. Surrey and A. Holmes proposed the concept of a plasma neutralizer where the beam itself would ionize the background gas along its path \cite{surrey1}, called the beam-driven plasma neutralizer (BDPN). The beam deposits energy in the neutralizer mostly in the form of stripped fast electrons (272~eV for a 1~MeV D$^{-}$ beam) and fast electrons generated by the beam through background gas ionization (called Rudd electrons). These fast electrons cause secondary ionization. In order to achieve a plasma density and degree of ionization sufficient for enhanced neutralization the walls are lined with cusp magnets. In a follow up in 2019 I. Turner and A. Holmes \cite{turner1} gave a more detailed description of a possible design for the BDPN and improved Surrey's and Holmes' original zero-dimensional model. For an ITER-like NBI ion beam they predicted an achievable neutralization efficiency of up to $\approx 80\;\%$.\\
The model was further amended by Starnella et al. \cite{starnella1}, who introduced several previously overlooked loss mechanisms, which led to a reduction of the predicted achievable neutralization to $\approx 68\;\%$. This latest model calculates also temperature of the neutralizer gas besides the degree of ionization and neutralization efficiency. The main inputs are the neutralizer main dimensions, cusp field strength and magnet separation as well as filling pressure. The model is iterative and takes several CPU hours to converge.\\ 
The BDPN design assumed by Starnella features a box-like neutralizer which, unlike the gas neutralizer for ITER, has no internal channel-separating walls in order to reduce plasma losses. All walls are lined with permanent magnets in Halbach configuration and there is a magnetic field across the beam entrance and exit slits created by magnets that sit in bars between the slits. As the absence of internal walls increases the gas conductance, the entrance and exit bars are extended towards the outside in the shape of "fins" (Figure \ref{fig:DEMO_NBI}b) as a compensatory measure to restrict the gas conductance. 
This design is of a conceptual nature only and not optimized for the integration into an actual beamline. In this paper  we attempt to do this; to optimize the shape parameters of the BDPN together with those of the ion source and the residual ion dump (RID) in order to maximise the neutral yield of the entire beamline, i.e. the flux of injected neutral D per D$^{-}$ ion current extracted from the ion source, rather than the neutralisation yield of the BDPN as a stand-alone component. This requires us to consider 
\begin{enumerate}
	\item the geometrical transmission losses for neutral D by all beamline components (BLCs) along the entire beamline and its dependence on BLC geometry parameters;
	\item the dependence of the plasma density and degree of ionization in the BDPN on BDPN dimensions and gas density;
	\item the gas temperature in the BDPN and gas conductance out of the BDPN in order to determine the required neutralizer gas flow;
	\item the neutral gas profile depending on the gas flow and pumping speed and spatial pump distribution in order to calculate reionization losses;
	\item the species evolution of the extracted D$^{-}$ beams along the entire beamline. 
\end{enumerate}
In order to make this optimization a manageable task, we prescribe the general shape of the BLCs and leave only their dimensional parameters free for optimization within an allowed range. Besides being convenient for the exercise, such restrictions will also exist in practice, e.g. in the form of a maximum allowed beamline envelope or a maximum tolerable opening in the breeding blanket.  

\section{Problem setup}
\label{problem}
Our starting point is the current pre-conceptual design for DEMO NBI (Figure \ref{fig:DEMO_NBI}). We keep the focus point, i.e. the point where all beamlets intersect, in the blanket and the distance from the neutralizer to the focus point. However, we allow the neutralizer length to change and reposition the ion source accordingly. On the vacuum side, the present cross-section of the NBI vessel is kept.\\
One also needs to decide how the beamlet array is organized: having tall beam columns is preferable for reducing gas conductance between the fins of the BDPN and in the channels of the RID. Furthermore, a cross section that is not too different from a square is beneficial for the BDPN because it keeps the volume-to-loss-surface ratio high and number of entrance slits should be even for reasons of symmetry of the magnetic cusp field. Hence the original layout with 4 by 4 beamlet groups array, each one consisting of 5 by 16 beamlets, appears to be a reasonable choice that we keep as well. However, while the vertical spacing between beam groups can be fixed beforehand because all of them  pass through the same slits and channels, the horizontal spacing needs to be a free parameter to allow the surrounding component geometry to change.\\
We imposed a number of further external constraints: the maximum horizontal dimension of the ion source is set to 1.5~m and the maximum allowed length of the neutralizer is 4~m.
The chosen work flow is as follows:

\begin{itemize}
	\item First we find the relations that determine the values of the BDPN and RID dimension parameters  necessary to achieve a set value for the geometrical transmission losses. Then we map the geometry parameter space for allowed solutions that are compatible with all external constraints, such as the size constraints mentioned above or a minimum allowed wall thickness.
	\item For every allowed point in the discretized parameter space from above the plasma density and neutral gas temperature in the BDPN is calculated using a parametrized scaling obtained with Starnella's model \cite{starnella1}. This calculation is carried out for various gas flows into the neutralizer. For each of these calculations the gas density distribution is calculated using a parametrized geometry in an FEM code. Said gas distribution can be used to solve numerically the system of equations describing the evolution of the D$^{-}$ beam through background gas collisions, and giving us the neutral beam equivalent current injected into the torus. 
	\item The geometry--gas-flow combination that gives the maximum neutralized beam fraction is the desired solution.
	\item Particle tracing simulation through the whole NBI beamline are then performed in order to calculate the resulting heat loads on the BLCs and to double-check the beam evolution and the transmission calculation.
\end{itemize}

\subsection{Parametric geometry description}
\label{paramgeom}
\begin{figure}
	\centering
	\includegraphics[width=0.9\columnwidth]{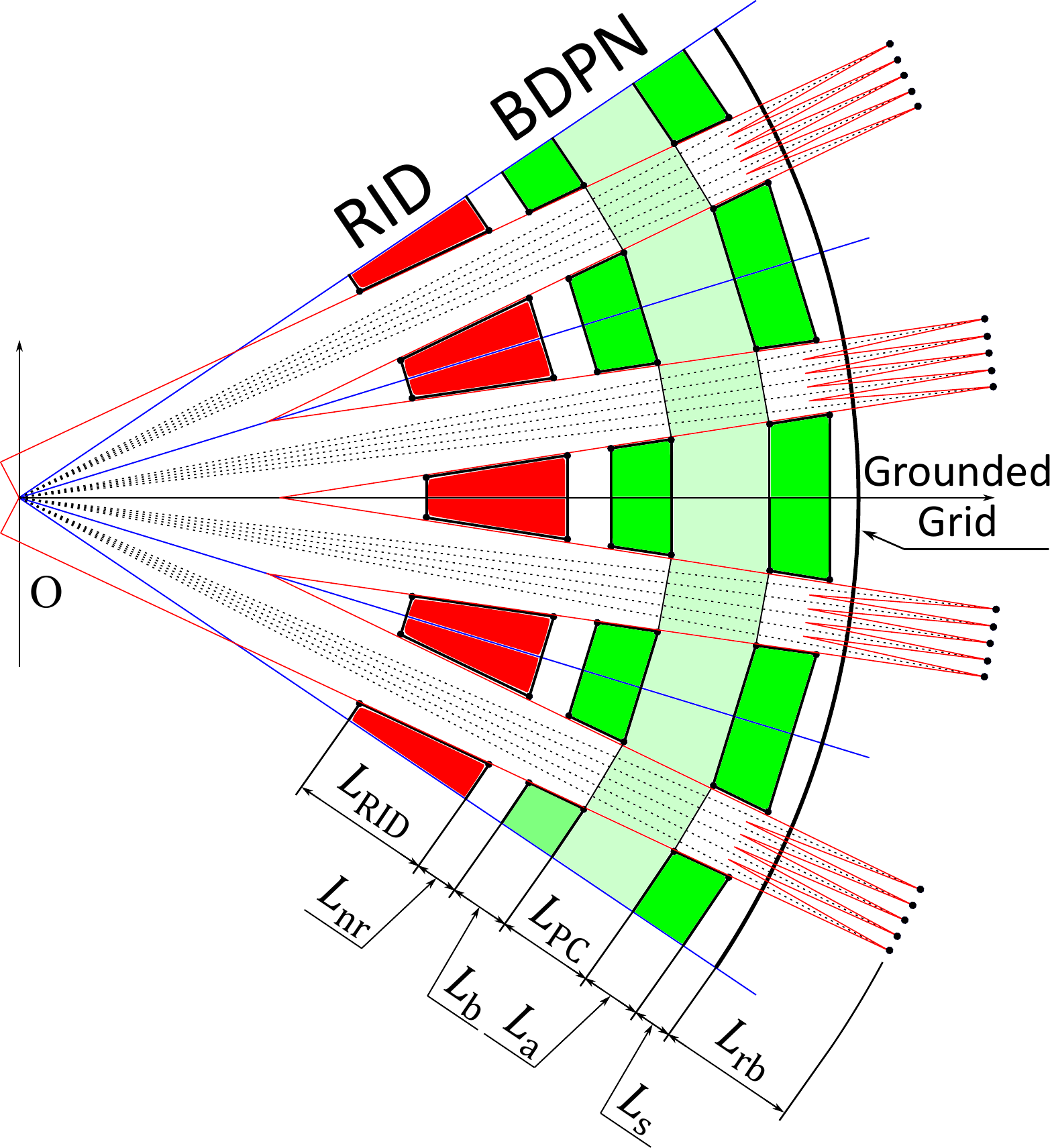}
	\caption{Geometry sketch of a horizontal section of the NBI BLCs in this parametric model.}
	\label{fig:NBI_sketch}
\end{figure}

Figure~\ref{fig:NBI_sketch} shows a schematic of the beamline geometry as a horizontal cross section. The grounded grid position is on circle on the right. The individual beamlets are described as starting from point sources on the right and expanding with a divergence angle $\epsilon$ as they travel towards the left. As in reality the beamlets have a finite size already on extraction, the point source is displaced some distance behind the extraction grid position. All beamlets are focused to a common point $O$ on the left where they intersect both horizontally and vertically in order to minimize the beam width in this position. Obviously, the beam width in point $O$ is then determined by the divergence of a single beamlet. In between the grounded grid and the focus point are the two beamline components of interest. The first in beam direction is the BDPN, characterized by an entrance fin length $L_\mathrm{a}$, an inner length $L_\mathrm{PC}$ (Plasma Chamber length) and an exit fin length $L_\mathrm{b}$. The second component is the RID with a length $L_\mathrm{RID}$ in a distance $L_\mathrm{nr}$ from the neutralizer. For simplicity it is assumed that all beamlet starting points as well as the BDPN and RID entrance and exit surfaces lie on spheres around point $O$. The neutralizer and RID are subdivided in four equal ``sectors'', one for each beamlet group, so that only one needs be determined and the others are obtained by rotation. The first task is to find an algorithm that determines the lengths and widths of the channels in the BDPN and RID that guarantees a requested geometrical transmission while satisfying all external constraints. It must be noted that Fig.~\ref{fig:NBI_sketch} is only a schematic in which the angles and hence the width-to-length ratio are strongly exaggerated.

\subsubsection{Beam model description}
Each beamlet is described by a Gaussian power distribution with a $1/e$ divergence $\epsilon$, which is defined as the angle between the main beam axis and the locations where the power density function is reduced by a factor of $1/e$. If the beamlet is aligned with the $z$ axis and starts from a point source the normalized power density distribution is defined as
\begin{equation}
	F(x,y,z) = \frac{1}{2 \pi \sigma^2(z)} \exp\left( -\frac{x^2+y^2}{2\sigma^2(z)}\right) ,
\end{equation}
where
\begin{equation}
	\sigma(z) = \frac{z \tan \epsilon}{\sqrt{2}}.
\end{equation}
Experimental NNBI beams however are better described by the superposition of two Gaussian distributions $F_{\mathrm{c}}$ and $F_{\mathrm{h}}$ sharing the total power of the beamlet at a given ratio. The part of the beam with the smaller divergence $\epsilon_\mathrm{c}$ is called \textit{core} and the remaining part with divergence $\epsilon_\mathrm{c}$ is called \textit{halo}. If $f_{\mathrm{c}}$ and $f_{\mathrm{h}} = 1-f_{\mathrm{c}}$ are the respective beam fractions for core and halo each beamlet can be described as
\begin{equation}
	\label{eq:total_beam}
	F_{\mathrm{b}}(x,y,z) = f_{\mathrm{c}} F_{\mathrm{c}}(x,y,z) + f_{\mathrm{h}} F_{\mathrm{h}}(x,y,z).
\end{equation}
For our calculations we assume $f_{\mathrm{c}} = 0.85$, $f_{\mathrm{h}} = 0.15$, $\epsilon_\mathrm{c} = 7\,\mathrm{mrad}$ and $\epsilon_\mathrm{h} = 15\,\mathrm{mrad}$, as specified for the ITER NBI. \\
The transmission of the entire beamline is ultimately determined by the component most limiting in horizontal and vertical opening angle as seen from the ion source, which is the connecting duct between the NBI vacuum vessel and the tokamak: the port size and duct must be small enough to fit between the toroidal field coils, and as small as possible to cut out as little as possible from the breeding blanket, limit the escaping neutron flux and the gas flow between NBI and torus.\\
As all beamlets are being focused from a long distance to a single point in the middle of the port to reduce the exit beam cross-section as much as possible, it is reasonable to equate the full beam transmission to the one of a single, centred beamlet expanding towards the exit; this means that Equation \ref{eq:total_beam} can be used to analytically estimate the transmission. For a square port (particular case of a rectangular port) of half-width $s$ and at a distance from the point source $L$, the normalized transmitted power is equal to
\begin{equation}
	\label{eq:transm_duct}
	p(s,L) = f_{\mathrm{c}} \erf^2\left(\frac{\alpha}{\tan\epsilon_{\mathrm{c}}}\right)+f_{\mathrm{h}} \erf^2\left(\frac{\alpha}{\tan\epsilon_{\mathrm{h}}}\right),
\end{equation}
where $\alpha = s/L$. We adopt a designer's approach and choose the size of the duct according to a desired $p_{\mathrm{t}}$. This is done by numerically inverting Equation \ref{eq:transm_duct} so that at a given distance $L$, the needed half-width $s_{\mathrm{t}}$ is obtained:
\begin{equation}
	\label{eq:slitpower}
	s_{\mathrm{t}} = s\left(p_{\mathrm{t}},L\right) = p^{-1}\left(p_{\mathrm{t}},L\right) .
\end{equation}
However, attention must be paid, since for the reasons mentioned above all apertures in the breeding blanket must be as small as possible.\\
This desired value $p_{\mathrm{t}}$ is the driving factor of design, and to make sure that it is respected, none of the components between source and duct exit may scrape more from the beams: in other words, a minimum buffer distance derived from Equation \ref{eq:slitpower} (varying as the beam expands along its sight line) from all beamlets' axes must be kept as they travel through the BLCs. However, since the BLCs design consists of very tall but thin channels to reduce gas conductance through them, usually the vertical ($y$) direction is not an issue for the components before the duct; this means that the BLCs can be designed in 2D.\\
This decision however affects the $p_{t}$, which would effectively increase, given that the duct appears no longer limited vertically; to this purpose, we distinguish between a $p_{t,\mathrm{tot}}$ and $p_{t,\mathrm{hor}}$, where $p_{t,\mathrm{tot}}$ represents the effective total power transmission in a three-dimensional duct, while $p_{t,\mathrm{hor}}$ represents the same quantity but restricted to the horizontal plane, which is the one that will be used in the design phase. The two are directly related, and for convenience we will be using only the latter; whenever $p_{t}$ is mentioned, we mean $p_{t,\mathrm{hor}}$.\\
In this approximation beamlets start from point sources, while having in reality a finite size ($r_{\mathrm{b}}=$ 7~mm): a backward displacement $L_{\mathrm{rb}}$ of the beamlet point sources with respect to the Grounded Grid mentioned above and shown in Fig.~\ref{fig:NBI_sketch} obtained inverting Equation \ref{eq:slitpower} is necessary, and in this case is approx. 0.6~m. $L_{\mathrm{rb}}$ must be added to all other distance values when determining slit widths.\\

\subsubsection{Geometric procedure}
To draw the delimiting surfaces of the BLC channels we first have to find the delimiting edge that is farthest downstream. For every beamlet this is the downstream edge of the RID channel wall that the beamlet is closest to. It is hence also sufficient to consider only the horizontally outermost beamlets for the respective RID channel edge.  
\begin{figure}
	\centering
	\includegraphics[width=0.9\columnwidth]{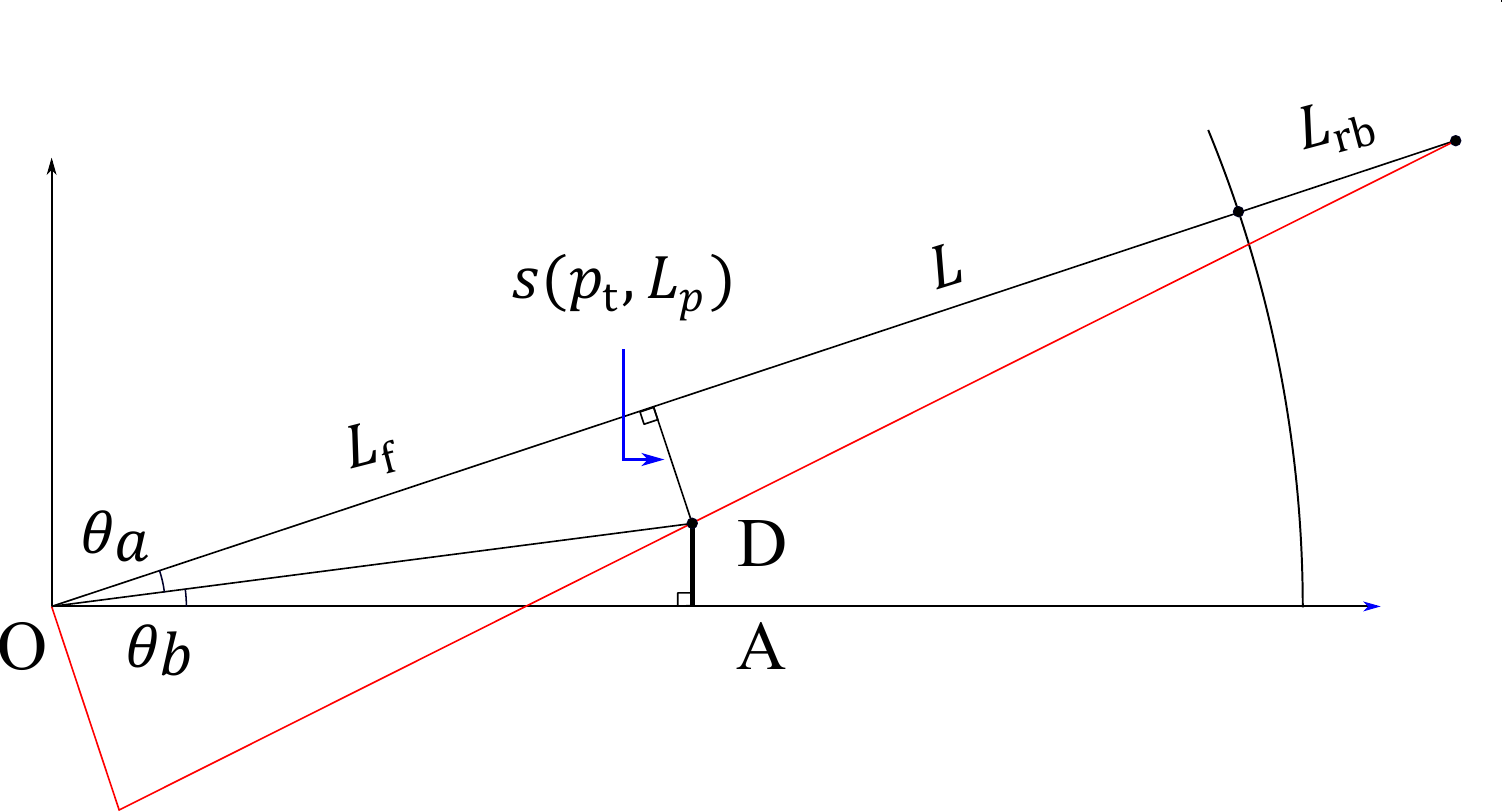}
	\caption{Geometry sketch for determining the initial point of the beam-fitting configuration.}
	\label{fig:bfc_first}
\end{figure}
By having the independent variable $L$ defined as the sum of the BDPN and RID lengths along the closest beamlet line of sight and postulating a desired $p_{\mathrm{t}}$, we can determine the necessary wall clearance $s$ of the outermost beamlet of a channel at the distance $L_{p} = L+L_{\mathrm{rb}}$, while the segment $\overline{AD}$ is half of the allowed minimum thickness of the RID plate between two channels. Once the remaining distance to the focus point $L_{\mathrm{f}}$ is fixed, the other quantities can be obtained as
\begin{align}
	\theta_{a} & = \tan^{-1}\frac{s\left(p_{\mathrm{t}},L_{p}\right)}{L_{\mathrm{f}}} \; ,\\ 
	\theta_{b} & = \sin^{-1}\frac{\overline{AD}}{\sqrt{L^{2}_{\mathrm{f}}+s^{2}\left(p_{\mathrm{t}},L_{p}\right)}}  \; ,\\
	\overline{AO} & = \sqrt{L^{2}_{\mathrm{f}}+s^{2}\left(p_{\mathrm{t}},L_{p}\right)-AD^{2}}.
\end{align}
The value $\theta_{t}=\theta_{a}+\theta_{b}$ is important since it describes the inclination of the first beam with respect to the initial horizontal axis; the first point of the configuration is then individuated by $D = \left(AO,AD\right)$.\\

\begin{figure}
	\centering
	\includegraphics[width=1\columnwidth]{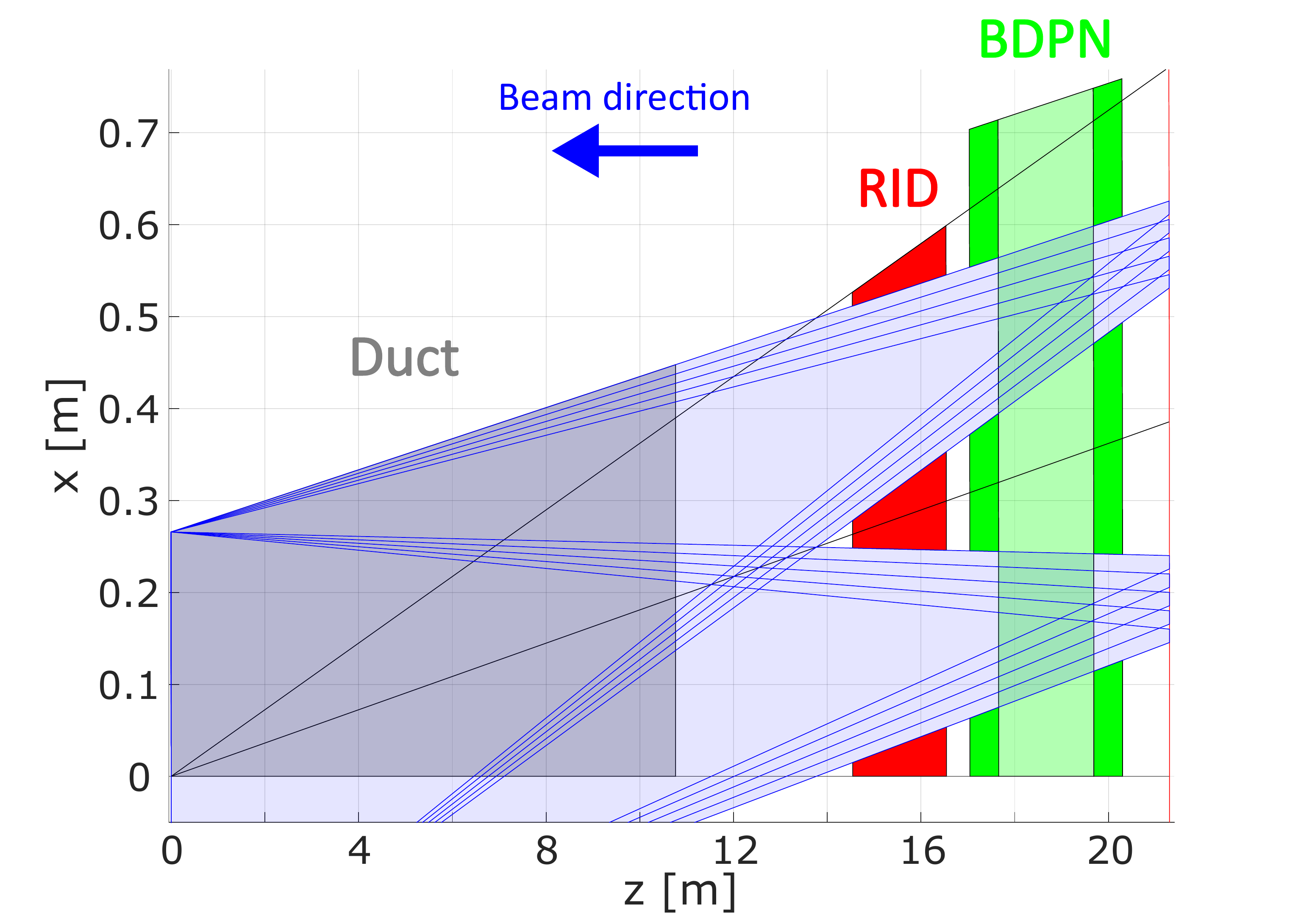}
	\caption{Plot of the geometry section of a possible beam-fitting solution for the BDPN. Not to scale.}
	\label{fig:bfc_result}
\end{figure}

All other delimiting points along the same channel wall in the RID and BDPN can be easily obtained because they lie on the straight line between the beamlet origin and $D$. The contour of the other side of the channel can be found by mirroring at the center beamlet axis and for other beamlet groups, i.e. other sectors, by rotation around point $O$. An example of this procedure applied to the whole NBI is shown in Figure \ref{fig:bfc_result}, while Table \ref{tab:variables} summarizes which are free and fixed quantities within our geometry description.\\ 
\begin{table}[h]
	\centering
	\begin{subtable}[h]{0.9\columnwidth}
		\centering
		\begin{tabular}{|l|c|}
			\hline
			\multicolumn{2}{|c|}{Free quantities} \\
			\hline
			Variable & Description \\
			\hline			
			$L_{\mathrm{PC}}$	& Plasma Chamber length\\
			$L_{\mathrm{a}}$	& Entrance fin length\\
			$L_{\mathrm{b}}$	& Exit fin length (same as $L_{\mathrm{a}}$)\\
			\hline
		\end{tabular}
	\end{subtable}
	\\
	\bigskip
	\begin{subtable}[h]{0.9\columnwidth}
		\centering
		\begin{tabular}{|l|c|}
		\hline	
		\multicolumn{2}{|c|}{Fixed quantities} \\
		\hline
		Variable & Description \\
		\hline
		$p_{\mathrm{t}}$    & Power transmitted at duct \\
		$2\cdot\overline{AD}$	& Minimum allowed RID width \\
		$L_{\mathrm{s}}$	& Distance GG to Neutralizer \\
		$L_{\mathrm{nr}}$	& Distance Neutralizer to RID \\
		$L_{\mathrm{RID}}$ 	& RID length \\
		$L_{\mathrm{f}}$	& Remaining distance to focus \\
		\hline
		\end{tabular}
	\end{subtable}
	\caption{Table summarizing the type of quantities used in describing the beam-fitting geometry.}
	\label{tab:variables}
\end{table}

In order to discriminate between feasible and non-feasible solutions, the resulting total neutralizer width is  compared to the maximum width allowed -- which, given the shape of the beam, is the most stringent boundary condition, since it affects directly the size of the necessary beam source -- and marked accordingly in the parameter space.

\subsection{Parametric FEM model description}
\label{param_FEM}
\begin{figure}
	\centering
	\includegraphics[width=0.9\columnwidth]{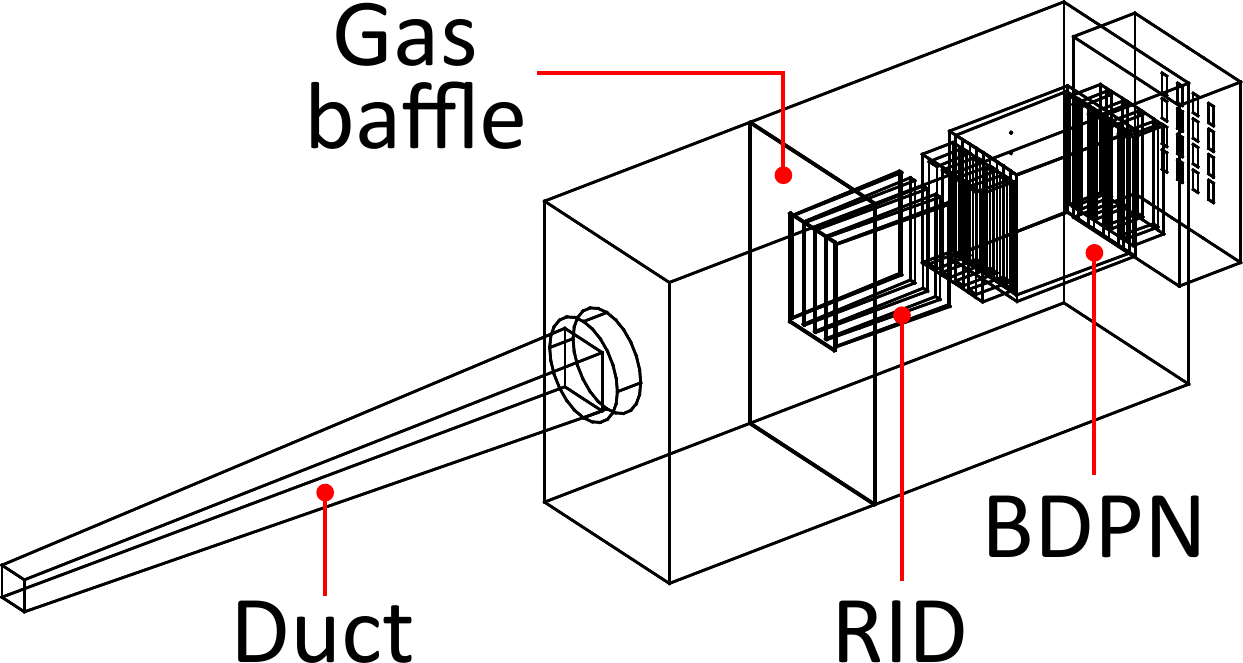}
	\caption{Geometry plot of the working DEMO NBI FEM model.}
	\label{fig:DEMO_model}
\end{figure}

The next step uses the information acquired during the geometry set exploration and builds for each one a 3D model of DEMO NBI based on the varying geometry of the BDPN. The gas flow in free molecular regime in this model is then solved by the Finite Element Method (FEM) commercial code COMSOL\textsuperscript{\textregistered} with the objective of obtaining the gas density distribution, while being controlled by a MATLAB\textsuperscript{\textregistered} script for post-processing. The formulation used by this code is an isothermal angular coefficient method, which treats gas flow in the same way as to radiation exchange: the code calculates the view factors between elements and models diffusion off of walls using a cosine law of reflection \cite{COMSOL_doc}.\\
The geometry, shown in Figure \ref{fig:DEMO_model}, is inspired by the latest DEMO NBI pre-conceptual design \cite{mqtran1} and tries to capture the main features of the system while simplifying it as much as possible to shorten computation time. The setup of the gas sources and sinks is also based on earlier reports on vacuum analysis in DEMO NBI (from \cite{xueli1}); these are: 
\begin{itemize}
	\item	The gas inflow into the neutralizer, $\Phi$, which is varied between simulations, is distributed among 5 gas inlets inside the plasma chamber.  
	\item 	The gas immission from the source through the accelerator, is set at $Q_{\mathrm{s}} =$ 3 $\mathrm{Pa}\;\mathrm{m}^{3}\;\mathrm{s}^{-1}$.
	\item	The gas entering the NBI duct from the tokamak, set at a conservative $Q_{\mathrm{c}} = $ 2 $\mathrm{Pa}\;\mathrm{m}^{3}\;\mathrm{s}^{-1}$.
	\item	The vacuum system is modelled by “pumping walls” on the top and bottom parts of the vessel, with a tentative capture coefficient $C_{\mathrm{s}} = $ 0.1.
\end{itemize}
The formulation of the model is isothermal, but the code still allows for surfaces at different temperatures within. The neutral gas in the neutralizer is heated by the transiting beam and this can be taken into account by assigning to the inner walls of the neutralizer the same temperature of the gas. The temperature gradient effect can be emulated in the passage between plasma neutralizer and room temperature by setting the appropriate value of $T_{\mathrm{NEU}}$  on the inner side. \\
The explorative nature of the problem at hand would in principle require a separate vacuum simulation for each variation of the gas inflow $\Phi$ or neutralizer temperature $T_{\mathrm{NEU}}$, which becomes immediately unreasonably time-consuming. To avoid this, the idea is to represent the variation of inflow and temperature through an appropriate weighted interpolation: this approach takes the solutions at the extremes of pre-established exploration boundaries (in our case, the solutions at minimum and maximum allowed inflow, minimum (room) and maximum allowed temperature) calculated by COMSOL and combines them through appropriate weighting in MATLAB to obtain an intermediate solution, which in a linear problem (as it is in this case) is very close to the one that would have been obtained by direct model evaluation. The set boundaries are between 300 and 600~K for the neutral gas temperature, and between $4\times 10^{21}$ and $2\times 10^{23}$~$s^{-1}$ for the neutralizer particle inflow. These boundaries were selected by trial and error, while trying to capture the full picture of the BDPN response to different flows and temperatures.
If $\left( T_{\mathrm{m}},T_{\mathrm{M}}\right),\left( \Phi_{\mathrm{m}},\Phi_{\mathrm{M}}\right)$ are the set boundaries, the weights are defined as
\begin{align}
	w_{T_{\mathrm{M}}}(T) = & \sqrt{\frac{T-T_{\mathrm{m}}}{T_{\mathrm{M}}-T_{\mathrm{m}}}}, \; w_{\Phi_{\mathrm{M}}}(\Phi) = \frac{\Phi-\Phi_{\mathrm{m}}}{\Phi_{\mathrm{M}}-\Phi_{\mathrm{m}}} \nonumber,\\
	w_{T_{\mathrm{m}}}(T) = & \; 1-w_{T_{\mathrm{M}}}, \; w_{\Phi_{\mathrm{m}}}(\Phi) = 1-w_{\Phi_{\mathrm{M}}}.
\end{align}
If $G = \left\lbrace k = (i,j) | \: i\in \left\lbrace T_{\mathrm{m}},T_{\mathrm{M}}\right\rbrace ,j\in \left\lbrace \Phi_{\mathrm{m}},\Phi_{\mathrm{M}}\right\rbrace \right\rbrace $ are the possible binary combinations of the respective cases, the new interpolated density is then obtained as:
\begin{align}	
	w_{k}(\Phi,T) = & \; w_{i} \; w_{j}; \; k \in G  \; ,\\
	n(\Phi,T) = & \sum_{k \in G}^{|G|} n_{k} \cdot w_{k}.
\end{align}
This interpolation is the main component of the overall MATLAB exploration algorithm which applies the BDPN scaling and loops through the available inflow values looking for the maximum neutralized beam fraction.

\subsection{The BDPN scaling}
\label{BDPN_scal}
\begin{figure}
	\centering	
	\begin{subfigure}[b]{1\columnwidth}
		\centering
		\includegraphics[width=\textwidth]{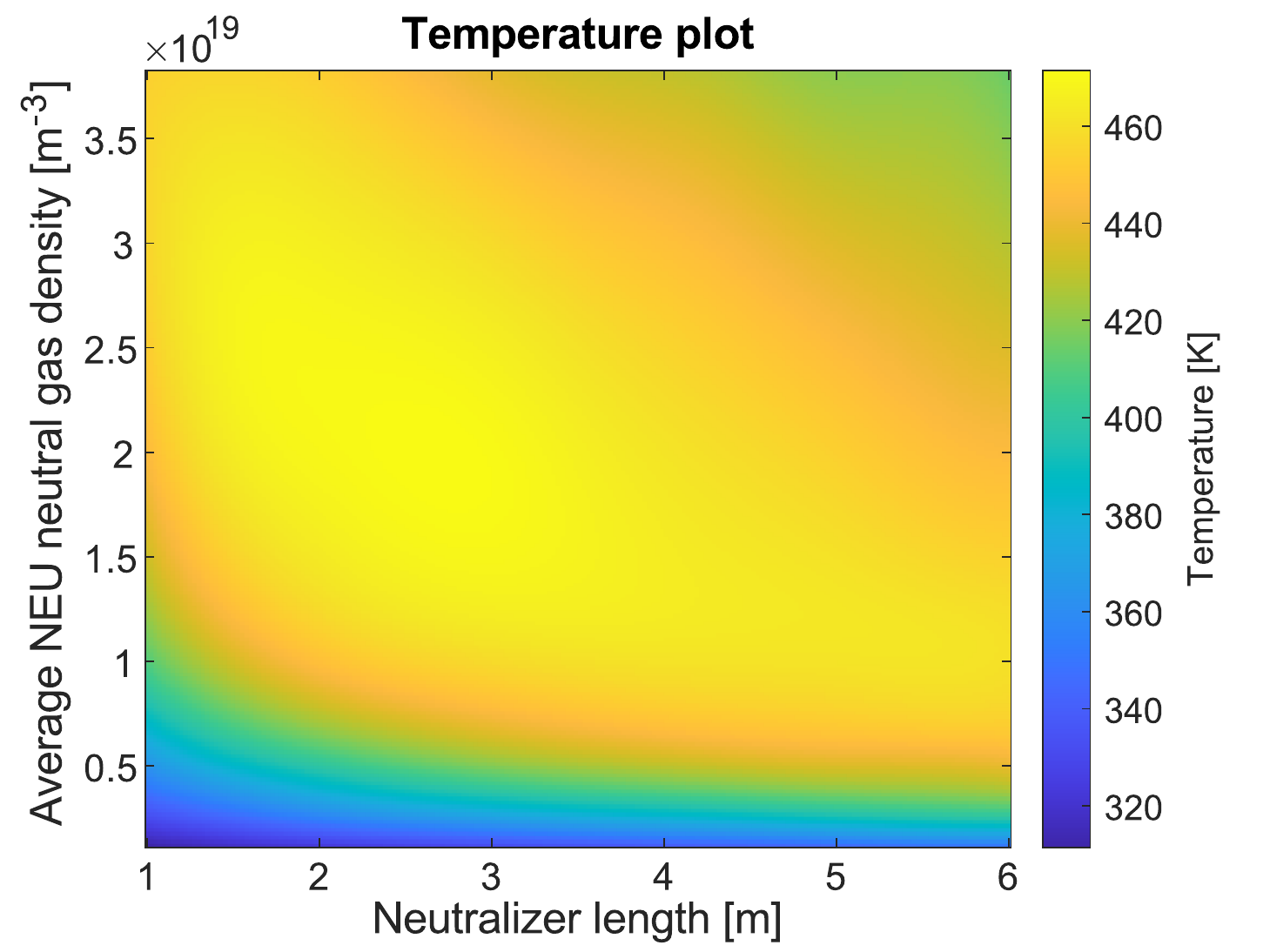}
		\caption{Temperature scaling for the BDPN. For a fixed neutralizer length, the denser the gas the more diluted the heating effect of the beam; for very low densities, the beam travels unaffected and does not heat the gas.}
		\label{fig:BDPN_temp}
	\end{subfigure}
	\bigskip
	\begin{subfigure}[b]{1\columnwidth}
		\centering
		\includegraphics[width=\textwidth]{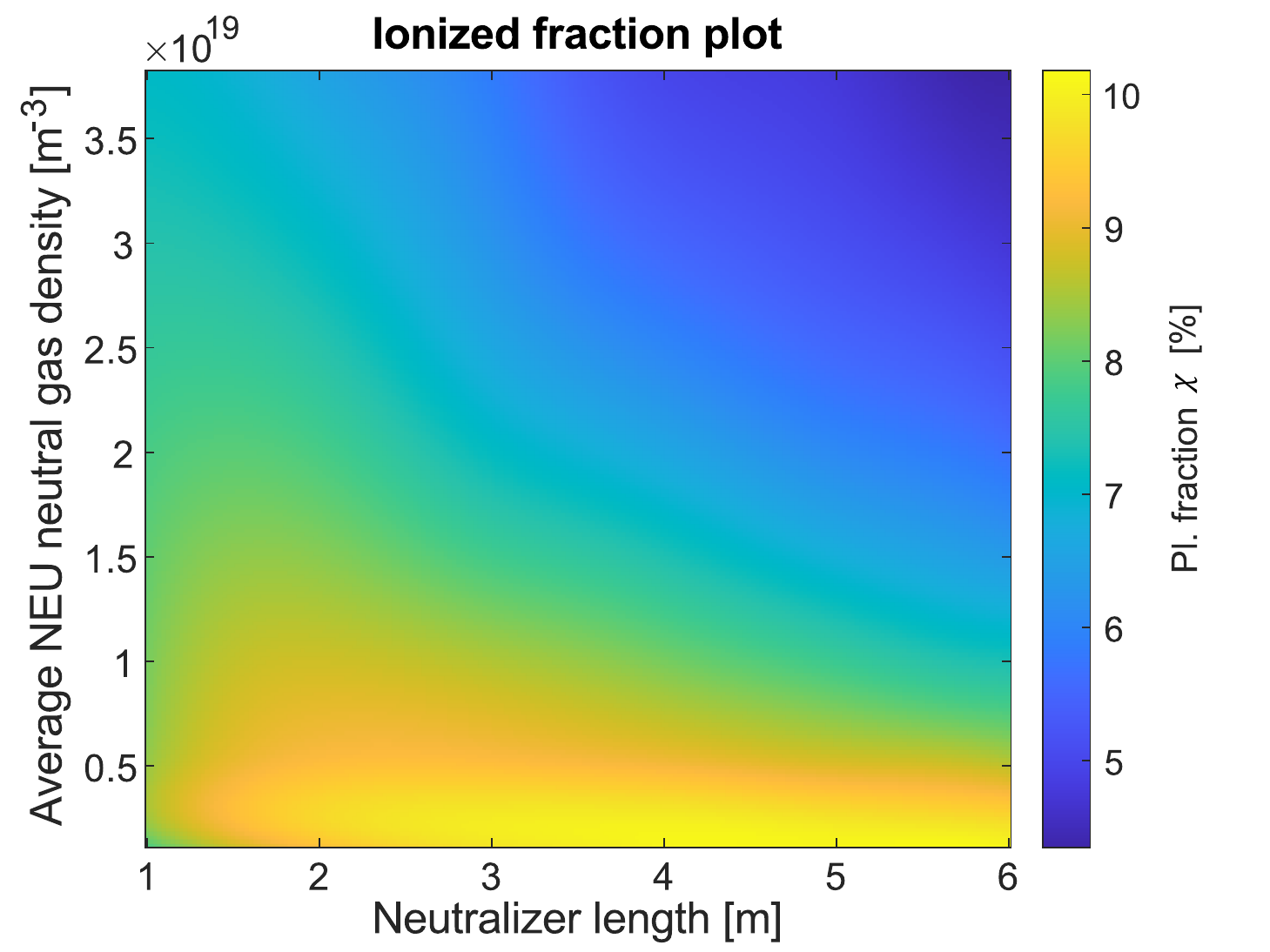}
		\caption{Ionization degree scaling for the BDPN. For a fixed neutralizer length, the denser the gas the more diluted the ionizing effect of the beam.}
		\label{fig:BDPN_ion}
	\end{subfigure}
	\caption{Plots of the BDPN scaling over their definition domain, at constant height and width.}
	\label{fig:BDPN_scal}
\end{figure}

In order to obtain the ionization degree and neutral gas temperature, the BDPN model of Starnella et al. \cite{starnella1} would have to be run for every point in the parameter space exploration described above. 
Starnella’s code is a zero-dimensional balance code that models the various populations of ions and electrons with their numerous interactions and loss channels (such as wall losses, impact dissociation, dissociative recombination, etc.) using as input the fast negative ion beam, BDPN geometry, and filling pressure, then yielding the resulting equilibrium temperature and expected ionization degree.
As one single BDPN model run takes several CPU hours, this is highly impracticable. To work around this problem the BDPN model was run for a reduced number of neutralizer lengths $L_\mathrm{PC}$, widths and heights as well as gas flows while maintaining all other numerous variables, such as the strength of cusp magnetic field, cusp spacing, etc., constant. The resulting data is stored as a multi-dimensional lookup table and the desired model outputs for a specific parameter combination can simply be obtained by interpolation. The dependences of gas temperature and ionization degree on neutralizer length as read from this table are shown in Figure~\ref{fig:BDPN_scal}.
\subsection{Exploration algorithm}
\label{exp_alg}
Once a point in geometry parameter space is deemed beam-optically viable and its weighted interpolation of the density is available from the lookup table, the exploration algorithm is engaged. Its purpose is to systematically scan the inflow parameter space available in the BDPN scaling lookup table while making sure that the solution is consistent regarding gas temperature and gas flow. This is done as follows:
\begin{itemize}
	\item The throughput loop is engaged: for each given value of $\Phi$, the average density inside the neutralizer is computed as a function of temperature between boundaries $T_{\mathrm{m}}$ and  $T_{\mathrm{M}}$. 
	\item This curve is intersected with the gas-temperature-vs-average-density curve from the BDPN scaling at the corresponding fixed neutralizer dimensions. The temperature at the intersection point is the working temperature $T_{\mathrm{NEU}}$ at which the calculation must continue.
	\item The final neutral density distribution is obtained by weighted interpolation using the given $\Phi$ and $T$ , while through the neutralizer average density and BDPN scaling the ionization degree $\chi$ is obtained.
	\item The neutral gas distribution and plasma density allow one to derive the beam fraction evolution through the NBI using cross-section data from literature (from Barnett \cite{barnett1} and Berkner \cite{berkner1}).
	\item	The process repeats for each allowed value of $\Phi$; the one that has the highest neutralized beam fraction at the exit of the NBI duct is deemed the optimum solution for the given geometric set. Other important quantities (such as corresponding specific inflow and temperature) are stored, ready to analyse the next geometry.
\end{itemize}
In this fashion it is possible to obtain a correlation between an optically viable geometry identified by the couple $\left(L_{\mathrm{a}},L_{\mathrm{PC}}\right)$ and respective quantities such as neutralized beam fraction at NBI exit, needed gas inflow, temperature; among which it is easy to look for the one with the highest neutral power.\\
As the entire algorithm could be hard to follow on text alone, Appendix \ref{appendice} contains a detailed flowchart of the process.

\subsection{Particle Tracing simulation}
\label{p_track}
The results of the previous step, namely the neutral gas density distribution, the plasma density distribution, and the geometry, are the inputs for the full-NBI particle tracing simulation in COMSOL; the outputs are the final transmitted neutral fraction to the plasma, as well as the heat loads due to particle loss on the various BLCs. The other boundary conditions are set as follows:
\begin{itemize}
	\item The particles' positions and directions are derived by starting from a 2D-Gaussian velocity distribution, whose width is dictated by the $1/e$ divergence chosen for the beam and peak related to the 1~MeV energy, and truncated at $ 3\sigma $ to avoid particles with very wide trajectories.
	\item Each beamlet is composed of a core and a halo fraction, with each of their guiding centres lying on the surface of a spherical GG and aimed towards the centre that coincides with the focus point.
	\item The total current assigned to the particles is 40~A, equal to the current extracted from the 1280 apertures (radius 7~mm) with an extracted current density of 254~A m$^{-2}$ and a supposed efficiency  of the accelerator of 0.8. This current is subdivided between core and halo components, with their respective fraction.
	\item The beam evolution is modelled through a collision algorithm using the same cross-sections as in \ref{exp_alg} to derive the reaction rate $\nu$.
	\item The only external fields considered are the one generated by the RID at 50~kV and the confinement magnetic field at the extremities of the BDPN; for now other sources (such as tokamak poloidal stray fields) have been ignored.
\end{itemize} 
In order to gauge the deflection effect of the end confinement magnets on the crossing charged particle beams, a tentative magnet configuration has been adopted with a feasible square cross-section of 20~mm of Sm-Co magnets.

\section{Results}
\label{result}

\subsection{Results of the BDPN model}
\begin{figure}
	\centering
	\includegraphics[width=1\columnwidth]{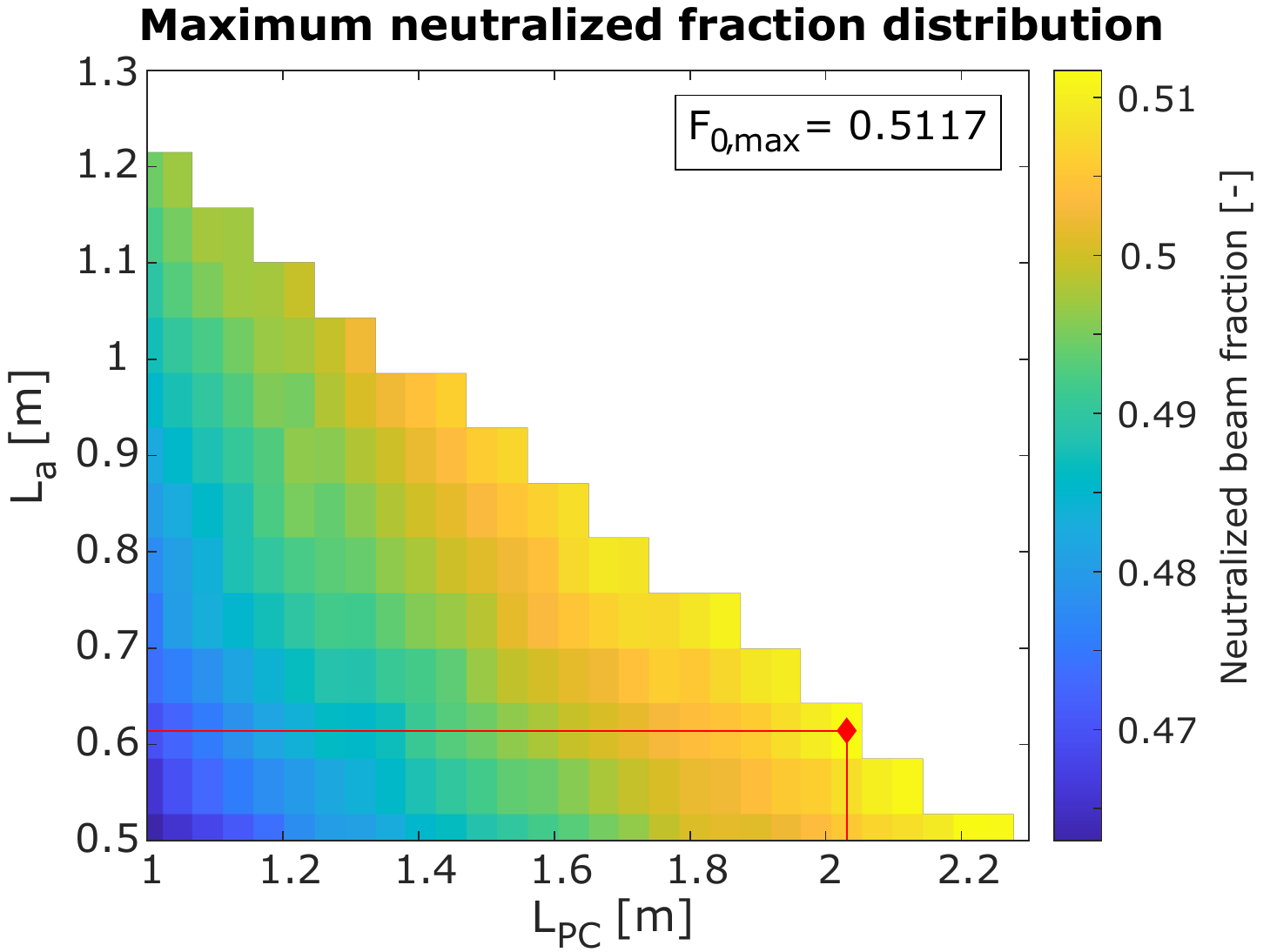}
	\caption{Map of the BDPN plasma chamber length ($L_\mathrm{PC}$) and fin length ($L_{\mathrm{a}} = L_{\mathrm{b}}$) parameter space showing the injected neutral fractions at optimal gas flow for each neutralizer configuration. The global maximum is indicated by a red diamond, while the white area on the top-right represents forbidden combinations. }
	\label{fig:F0_sol}
\end{figure}

\begin{figure}
	\centering
	\includegraphics[width=1\columnwidth]{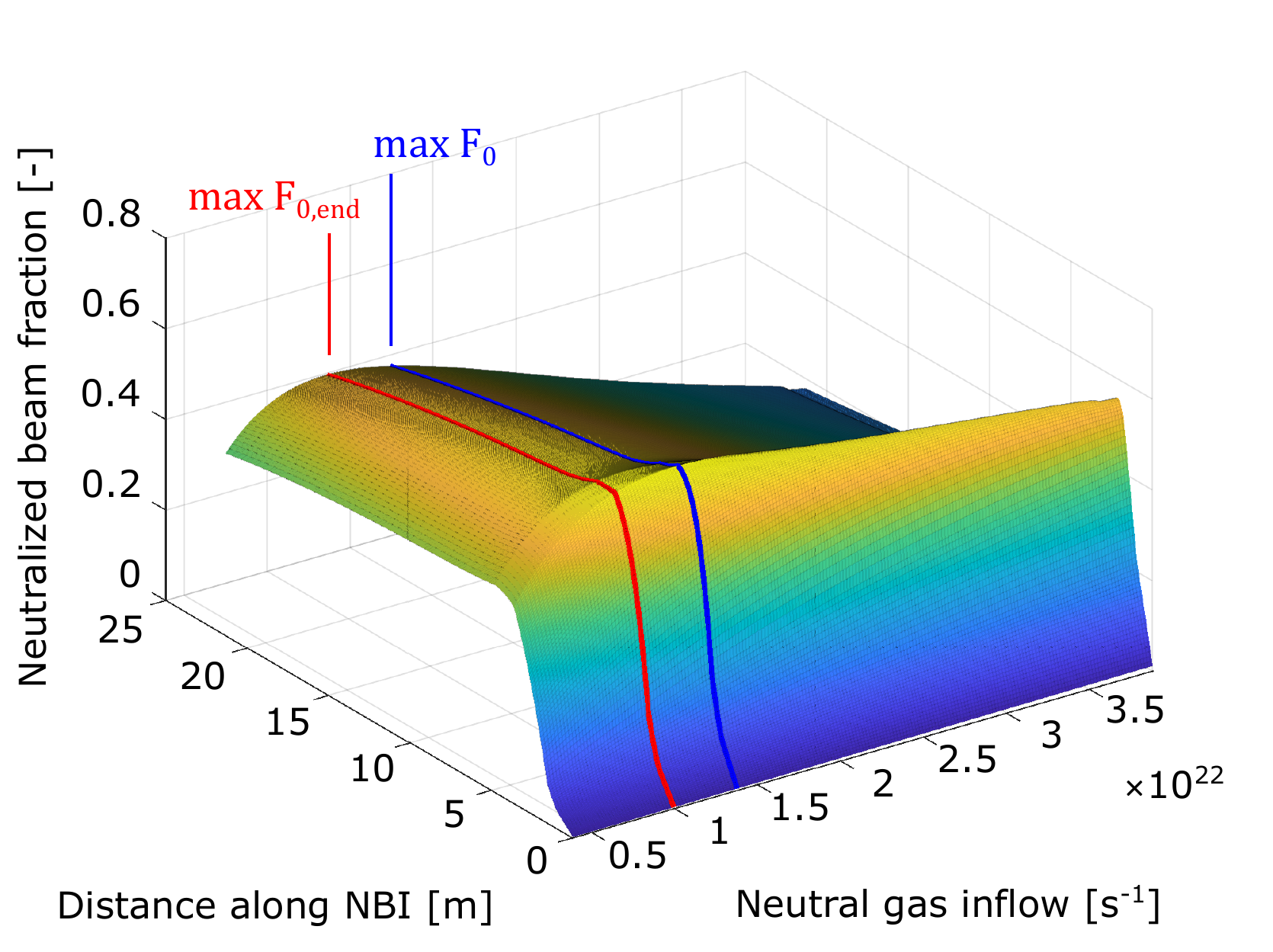}
	\caption{Plot of the evolution of the neutral fraction along the NBI (right to left) as a function of gas inflow. The red line represents the beam evolution that has the maximum neutral fraction at the NBI exit, while the blue line represents the one that reaches the maximum at the end of the neutralizer.}
	\label{fig:J_F0_sol}
\end{figure}

The parameter exploration yielded interesting results; by collecting each feasible combination of fin and neutralizer length with their respective maximized neutral fraction at the NBI exit in a plot, one can visualize where the absolute highest value is located (Figure \ref{fig:F0_sol}). It is important to specify that these numbers are not wall-plug efficiencies, but rather neutralization yields, and only part of the latter. An estimate of the wall-plug efficiency by using the available data will be given later in the paper.\\
The maximum value is 0.5117, located at $L_{\mathrm{PC}} =$ 2.03~m, $L_{\mathrm{a}} = L_{\mathrm{b}} =$ 0.614~m, with a gas inflow of $\Phi = 1.01\times 10^{22}~\mathrm{s}^{-1}$ (equivalent to 38 $\mathrm{Pa}\;\mathrm{m}^{3}\;\mathrm{s}^{-1}$ at 0\textdegree C). The triangular shape of the solution space is a consequence of the maximum width constraint, with the upper edge marking the points where the overall neutralizer length $L_{\mathrm{NEU}} = L_{\mathrm{a}}+L_{\mathrm{PC}}+L_{\mathrm{b}} = 2L_{\mathrm{a}}+L_{\mathrm{PC}}$ is at its achievable maximum.\\
Another aspect can be appreciated while looking at a plot showing the evolution of the neutral fraction along the beamline as a function of the gas input (Figure \ref{fig:J_F0_sol}).
Intuitively, when trying to maximize the neutral fraction, one would place the working point of the neutralizer as close to the available maximum as possible; however, the plot in Figure \ref{fig:J_F0_sol} shows a different picture:  the ideal beam evolution granting maximum neutral fraction at the beamline end, $\max F_{0,\mathrm{end}}$) is at lower neutralizer gas flow than that which achieves maximum neutralization ($\max F_{0}$) at the neutralizer exit. This is due to the increased gas requirement to obtain the maximum neutralizer performance, which in turn increases the density after the neutralizer and the reionization losses, losing more than what was gained.

\subsection{Comparison with a gas neutralizer}
\begin{figure}
	\centering
	\includegraphics[width=1\columnwidth]{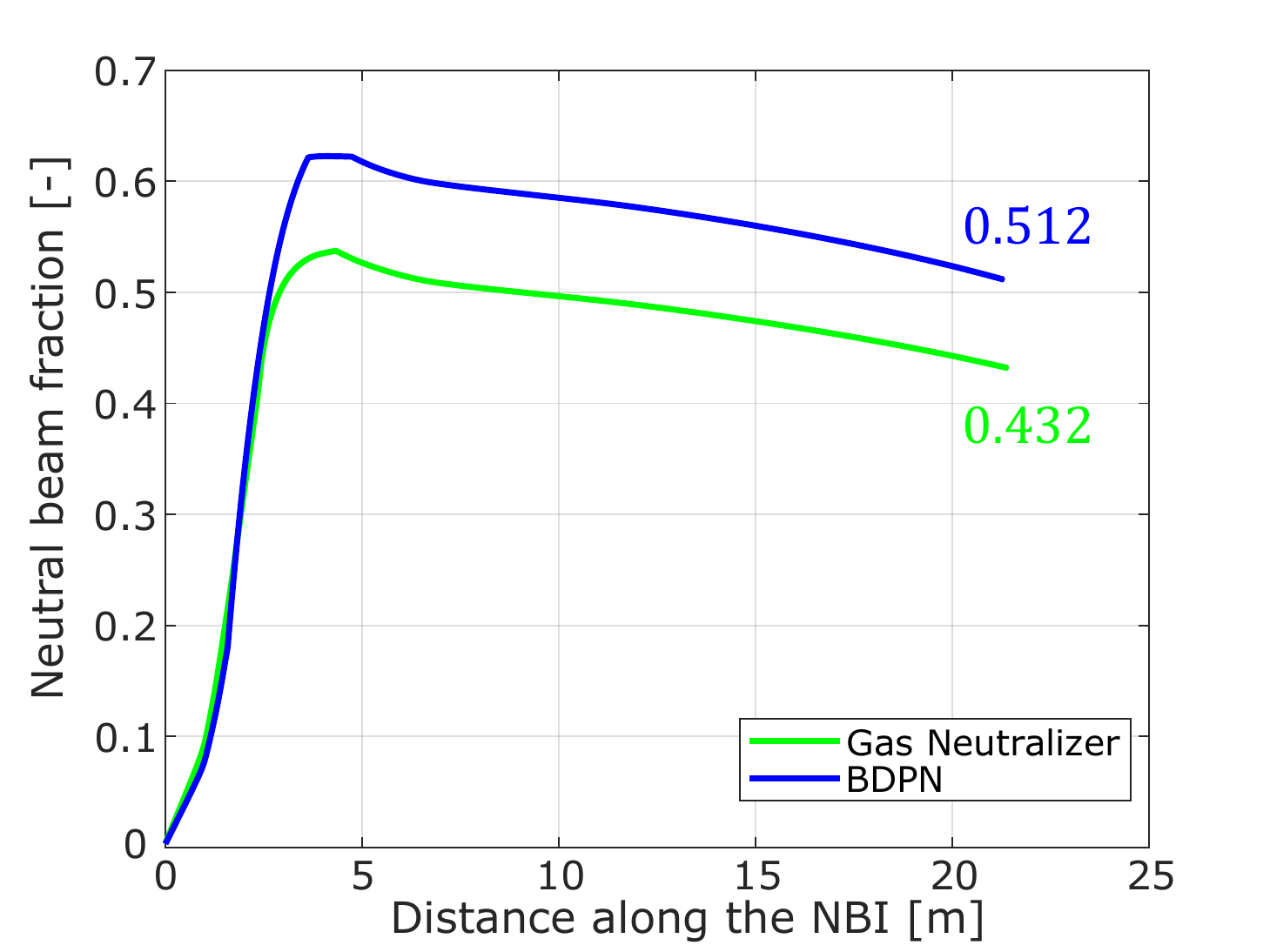}
	\caption{Comparison between GN (green) and BDPN (blue). The maximum neutral fraction at the exit of the NBI is shown.}
	\label{fig:gas_sol}
\end{figure}

\begin{figure}
	\centering
	\includegraphics[width=1\columnwidth]{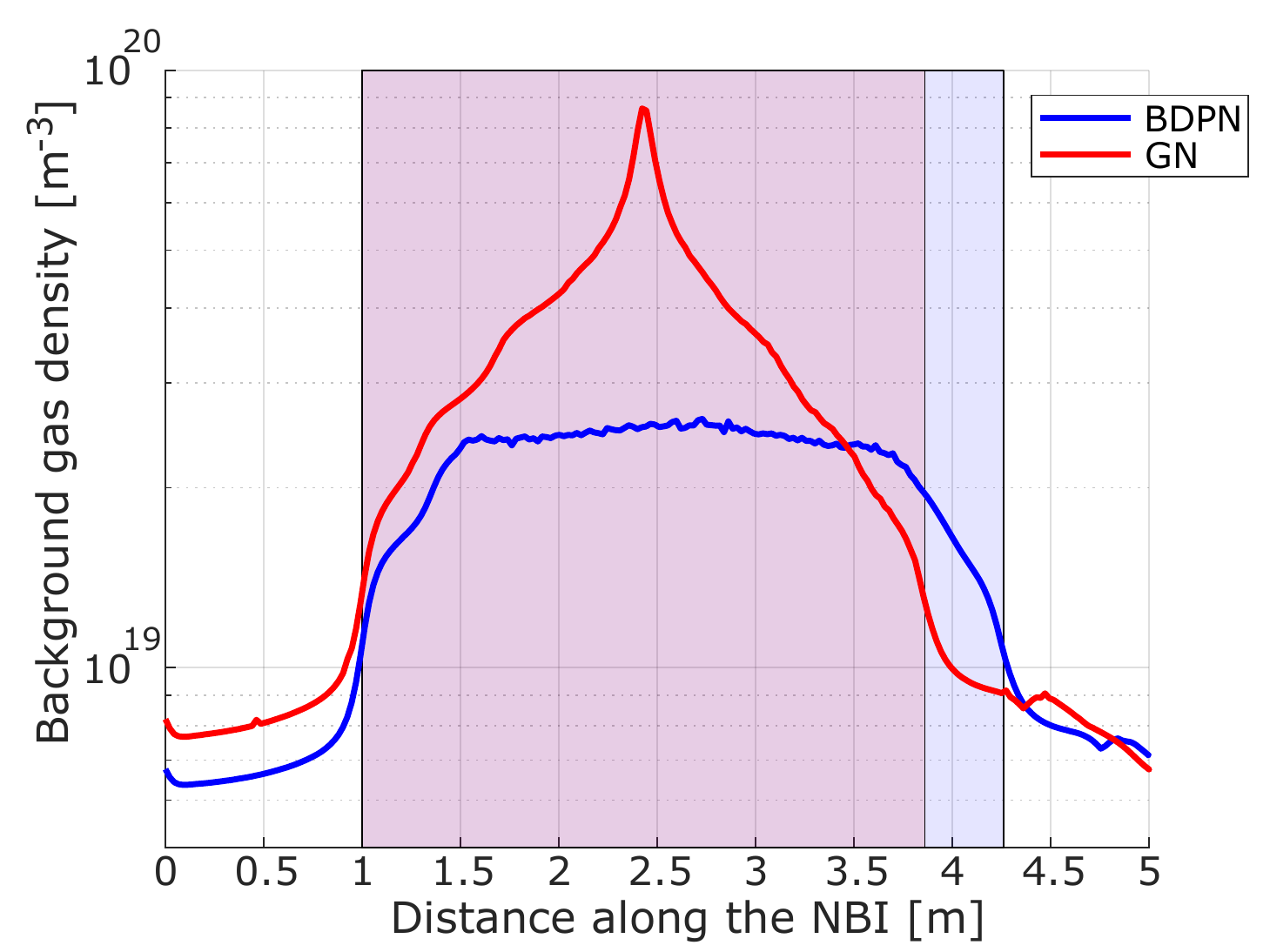}
	\caption{Comparison between the background gas distribution between GN and BDPN in the Neutralizer region. The outline of the Neutralizer in the two cases is reported.}
	\label{fig:dens_sol}
\end{figure}

Since the BDPN has been proposed as a replacement for the gas neutralizer (GN), it is important to compare the BDPN result in a meaningful way with a GN. Hence we compare it with an "equivalent" gas neutralizer (GN), where equivalent refers to:
\begin{itemize}
	\item Same maximum allowed encumbrance (length and width).
	\item Same beam quality and desired power fraction.
	\item Same beam-fitting design applied, equivalent to a BDPN but with a constant $L_{\mathrm{PC}} = 0$.
	\item Same maximization of neutral fraction by tuning the gas inflow; there are now however 3 gas inlets for each one of the 4 channels (for a total of 12 apertures).
\end{itemize}
The best neutral beam evolution in both cases are shown in the same plot to highlight the difference (Figure \ref{fig:gas_sol}), as well as the neutral gas distribution (Figure \ref{fig:dens_sol}).
The comparison shows that a GN achieves around 8 percentile points less than that its BDPN counterpart ($F_{0,\mathrm{end}} = $ 0.432), with a similar length of 2.86~m. The total inflow needed in this case is $\Phi = 1.09\times 10^{22}~\mathrm{s}^{-1}$ (41 $\mathrm{Pa}\;\mathrm{m}^{3}\;\mathrm{s}^{-1}$ at 0\textdegree C).\\
It is possible to derive an estimate of the NBI wall-plug efficiency in the two cases in a similar fashion as what has been done in \cite{mqtran1}: by substituting in the chain of efficiencies the new values from the results, the wall-plug efficiency is 0.30 for the GN and 0.34 for the BDPN (Table \ref{tab:DEMO_eff}).\\
\begin{table*}[t]
	\centering
	\begin{tabular}{|l|>{\centering}p{2 cm}|>{\centering}p{2.7 cm}|>{\centering\arraybackslash}p{2.7 cm}|}
			\cline{2-4}
			\multicolumn{1}{c|}{\multirow{2}{*}{ }} & Reference & Option 1 & Option 2 \\
			\cline{2-4}
			\multicolumn{1}{c|}{}& ITER HNB & 2022 DEMO NBI (with GN) & 2022 DEMO NBI (with BDPN)\\
			\hline 
			Extracted current density [A m$^{-2}$]		& 289 	& 251 	& 224	\\
			\hline
			Total extracted current [A]					& 57 	& 50 	& 44	\\
			\hline
			Nominal acceleration voltage [MV]			& 1 	& 1 	& 1		\\
			\hline
			Aux/extraction overall efficiency			& 0.9 	& 0.9 	& 0.9	\\
			\hline
			Gross power [MW]							& 63 	& 55 	& 49	\\
			\hline
			Stripping/halo current losses efficiency	& 0.7 	& 0.8 	& 0.8	\\
			\hline 
			Accelerated current [A]						& 40 	& 40 	& 35	\\
			\hline 
			Source/neutralizer transmission  			& 0.95 	& 0.95 	& 0.95	\\
			\hline
			Neutralizer efficiency						& 0.55 	& 0.54 	& 0.62	\\
			\hline
			Beam line/duct transmission 				& 0.8 	& 0.82 	& 0.8	\\
			\hline
			Estimated power to the plasma [MW]			& 16.67 & 16.67 & 16.67	\\
			\hline
			Injector overall efficiency					& 0.26 	& 0.30 	& 0.34	\\
			\hline
	\end{tabular}
	\caption{Main parameters and efficiencies for the different DEMO NBI options. The first column is from \cite{mqtran1}, while the other two incorporate the results obtained here.}
	\label{tab:DEMO_eff}
\end{table*}
\subsection{Results of the Particle Tracing model}
\begin{figure}
	\centering
	\includegraphics[width=0.95\columnwidth]{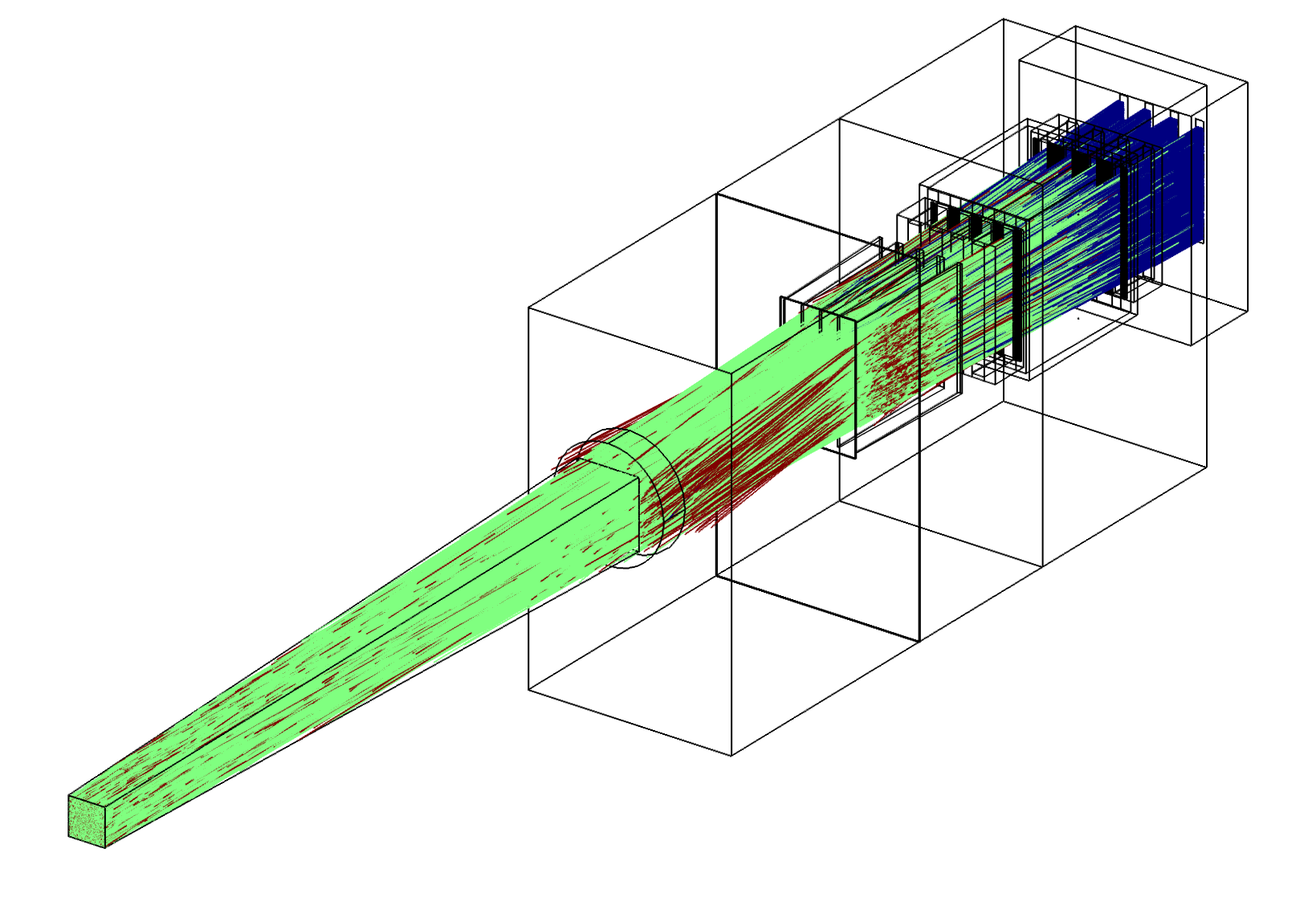}
	\caption{Trajectory plot of the particle model. Blue are the D$^{-}$ ions, green the D$^{0}$ neutrals, and in red the D$^{+}$ ions.}
	\label{fig:trajectory_plot}
\end{figure}

The Particle Tracing model has been run in order to determine the heat loads distribution, the power balance, and deliver a valuable cross-check to confirm the earlier results (Figure \ref{fig:trajectory_plot}).
To assess the effect of the end magnets on the beam ions, an otherwise identical simulation without magnets was also run. Furthermore, to cross-check if the beam-fitting design is able to transmit the desired power, another simulation without reactions and external fields was launched. The results are listed in Table \ref{tab:part_sol}.\\
\begin{table}[h]
	\centering
	\begin{subtable}[h]{0.9\columnwidth}
		\centering
		\begin{tabular}{| l |c  c|}
			\hline
			 & \multicolumn{1}{c|}{No Mag. field} & w/ Mag. field \\
			\hline
			Initial power	& \multicolumn{2}{c|}{40.04 MW}  \\
			Neutralizer 	& 0.671 MW	& 0.705 MW \\
			RID 			& 14.90 MW 	& 15.04 MW \\
			Exit (D$^{0}$)	& 19.30 MW 	& 17.29 MW \\
			\hline
			GG - plasma 	& 48.20\;\% & 43.18\;\% \\
			\hline
		\end{tabular}
	\end{subtable}
	\\
	\bigskip
	\begin{subtable}[h]{0.9\columnwidth}
		\centering
		\begin{tabular}{| l | c | c |}
			\hline
			  & Numeric & Analytic \\
			\hline
			Transmission & 91.92\;\% & 91.90\;\% \\
			\hline
		\end{tabular}
	\end{subtable}
	\caption{Results of the various particle simulations for the BDPN. Upper table from top to bottom: initial power launched at the grounded grid, power lost in the neutralizer, power dumped in the RID, power at the beamline exit into the torus, fraction of initial power that is injected; Lower table: comparison of geometrical transmission calculated by the analytic approach and numerical particle tracing without any magnetic or electric fields.}
	\label{tab:part_sol}
\end{table}

\begin{figure}
	\centering
	\includegraphics[width=1\columnwidth]{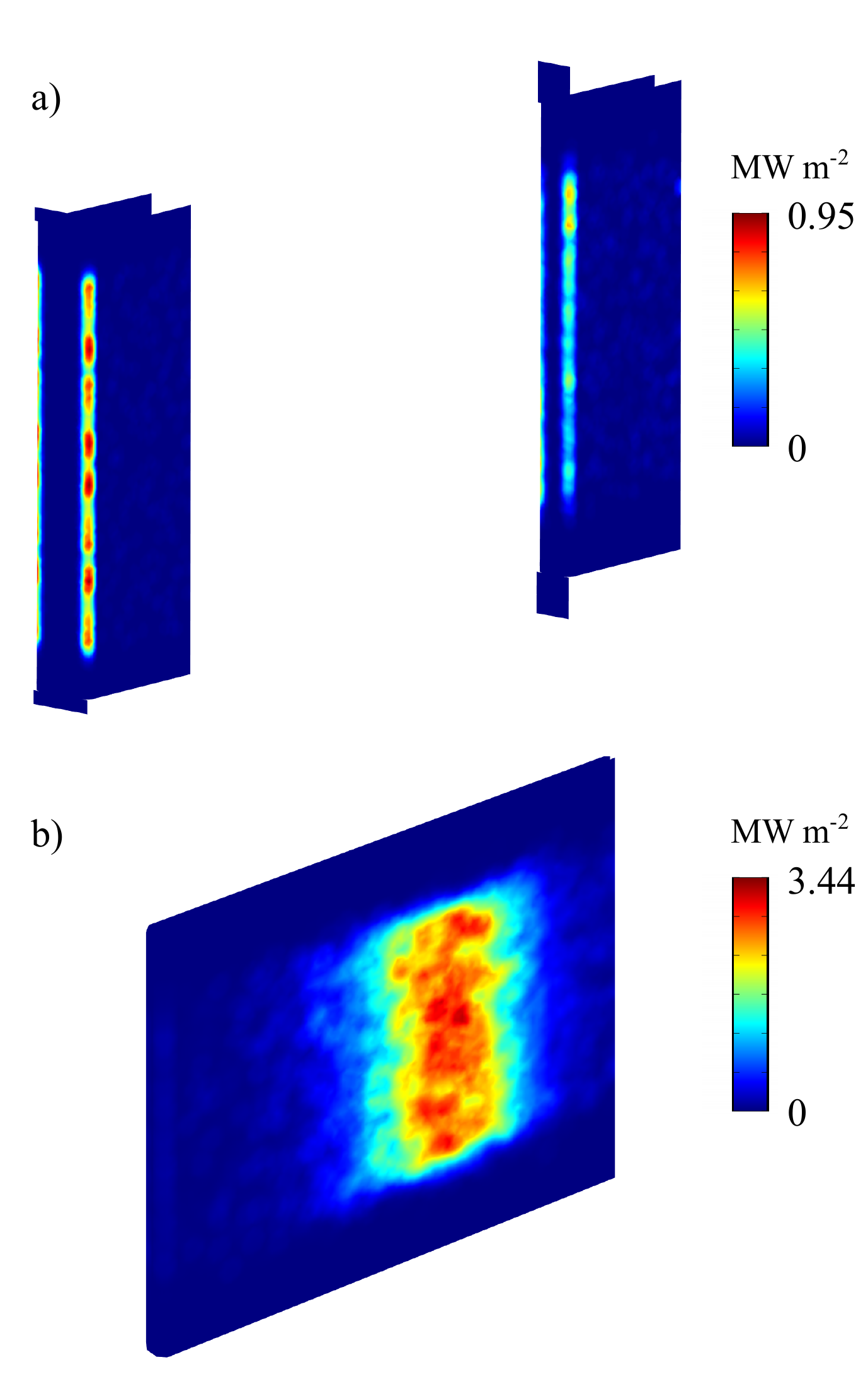}
	\caption{Heat load maps for the interested BLCs. Figure a) is the central set of fins of the neutralizer, while figure b) is the central plate of the RID.}
	\label{fig:heat_load_plot}
\end{figure}

The simulations show a very close match between the analytic estimation of transmission and the numeric result, confirming the premise of the beam-fitting design approach. Another important confirmation to get is on the neutral fraction achieved: in MATLAB this value is obtained through a system of differential equations, while in COMSOL the beam evolution is obtained through particle collision algorithms. To check this aspect, it is possible to divide the neutral fraction at the end of the NBI in the no magnet case (0.4820) by the numerical transmission (0.9192) to obtain the effective numerical neutral conversion fraction 0.524, to be compared with the original $F_{0,\mathrm{max}}$ of Figure \ref{fig:gas_sol} (0.512). The two values differ for merely about 3\;\%, a very good result given the degree of approximation.\\
As a consequence of the beam-fitting approach, the heat load power density is evenly distributed, with peaks around 0.95~MW m$^{-2}$ for the Neutralizer fins and 3.44~MW m$^{-2}$ for the RID panels (Figure \ref{fig:heat_load_plot}).\\
The data in Table \ref{tab:part_sol} underlines also the importance of choosing the appropriate magnet configuration at the ends of the BDPN: with the one used in the model, inserted as a first tentative, the deflection is enough to cause additional losses of 2~MW.\\
\subsection{Possibility for further improvement}
\begin{figure}
	\centering
	\includegraphics[width=1\columnwidth]{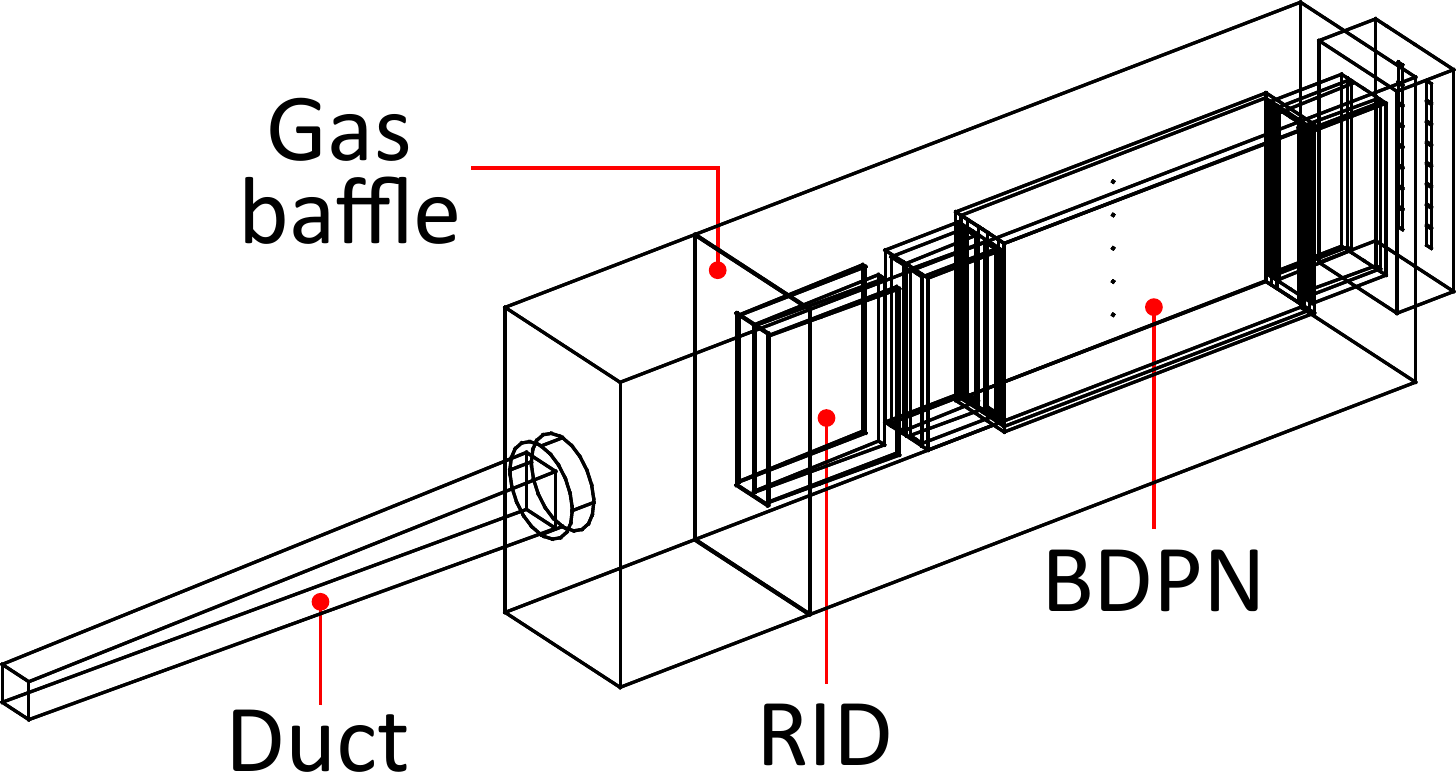}
	\caption{Geometry plot for the two-blade beam design.}
	\label{fig:BDPN_two_beam}
\end{figure}

The presented results constitute an optimization under the initial boundary conditions that have been set, and are therefore not a global optimum. The optimization exercise has also shown how different quantities influence each other. These trade-off relations can be used to imagine new possible geometric configurations where some constraints are more relaxed, and better performance can be achieved.\\
For example, we examined also a two-blade beam design, obtained by reshaping the beam group array from the 4 by 4 into a 2 horizontal by 8 vertical beamlet groups configuration, similarly to a previously considered DEMO NBI design \cite{sonato1}. The height of the vessel and BLCs is then appropriately increased to fit the new beam shape, resulting in the geometry of Figure \ref{fig:BDPN_two_beam}. The advantage of such a configuration is that the width of the BLCs is greatly reduced, and almost constant as the beamline gets longer: this is due to the expansion boundary of the side beamlets being almost completely compensated by the focusing towards the center. Obviously this cannot go to extreme lengths, since also the blanket aperture must scale accordingly, however it allows for a significantly longer neutralizer.\\
The BDPN scaling was extended in order to cover neutralizer lengths up to 6~m, and the optimization routine run again.\\
The highest value of neutralization efficiency of 0.56 (5 percentile points more than the four-blade beam) was found on the extreme upper right corner of its parameter space plot (obtained in the same way as Figure \ref{fig:F0_sol}), which means that the space size chosen this time (up to $\max(L_{\mathrm{a}},L_{\mathrm{PC}}) = ( \text{1.5~m} , \text{6~m})$, as large as the scaling currently allows) was still too small to contain the global maximum. Still, this trial proves that different approaches to the choice of geometry may lead to even more enhanced neutralization rates.

\section{Conclusions}
In this paper, for the first time, a complete NNBI beamline with a BDPN has been modelled and its parameters optimized, assuming realistic geometry and boundary conditions. Considering geometric transmission, the degree of ionization achieved in the plasma neutralizer as a function of its geometry and feed gas flow, the neutral gas distribution in the entire beamline and the beam species evolution along the whole beamline we calculate a neutral yield from the grounded grid of the ion source's accelerator to the beamline end, i.e. the injected D$^0$ flux divided by the accelerated D$^-$ flux, of about 48\;\% and a wall plug efficiency of the beamline of 0.34. This is about 6 percentile points more neutral fraction and about 4 percent more wall plug efficiency than what we calculate for a gas neutralizer, applying the same optimization procedure and boundary conditions.\\
The optimization exercise has highligted some more and some less intuitive tradeoffs. Obviously making the BDPN, and thus the whole beamline, longer means that the duct opening in the breeding blanket either has to be larger or leads to more geometrical losses. However, the combined optimization of neutralizer geometry and gas flow in order to obtain an optimal neutral flux at the beamline end favours a longer neutralizer. Interestingly, for a given geometry the gas flow that results in the highest neutral fraction at the exit of the neutralizer is higher than the gas flow that creates the highest neutral fraction at the beamline end. This is due to the additional reionization losses downstream of the neutralizer caused by the additional gas, which overcompensate the increased neutralisation in the BDPN. Understanding the various trade-offs is indispensable in driving the future design effort. \\
Obviously, the values reported here are optimizations under a particular set of boundary conditions. These boundary conditions were guided by space envelope of the current NBI beamline design for the European DEMO, which has a gas neutralizer, but are to some extent arbitrary, as every one of them can be disputed. However, the optimization  framework developed here can be used to guide the design under any set of external boundary conditions once these are known.\\

\section{Acknowledgements}
This work has been carried out within the framework of the EUROfusion Consortium, funded by the European Union via the Euratom Research and Training Programme (Grant Agreement No 101052200 - EUROfusion). Views and opinions expressed are however those of the authors only and do not necessarily reflect those of the European Union or the European Commission. Neither the European Union nor the European Commission can be held responsible for them.

\newpage
\appendix
\onecolumn
\section{Flowchart of the exploration algorithm}
\label{appendice}
\begin{figure*}
	\centering
	\includegraphics[width=0.79\textwidth]{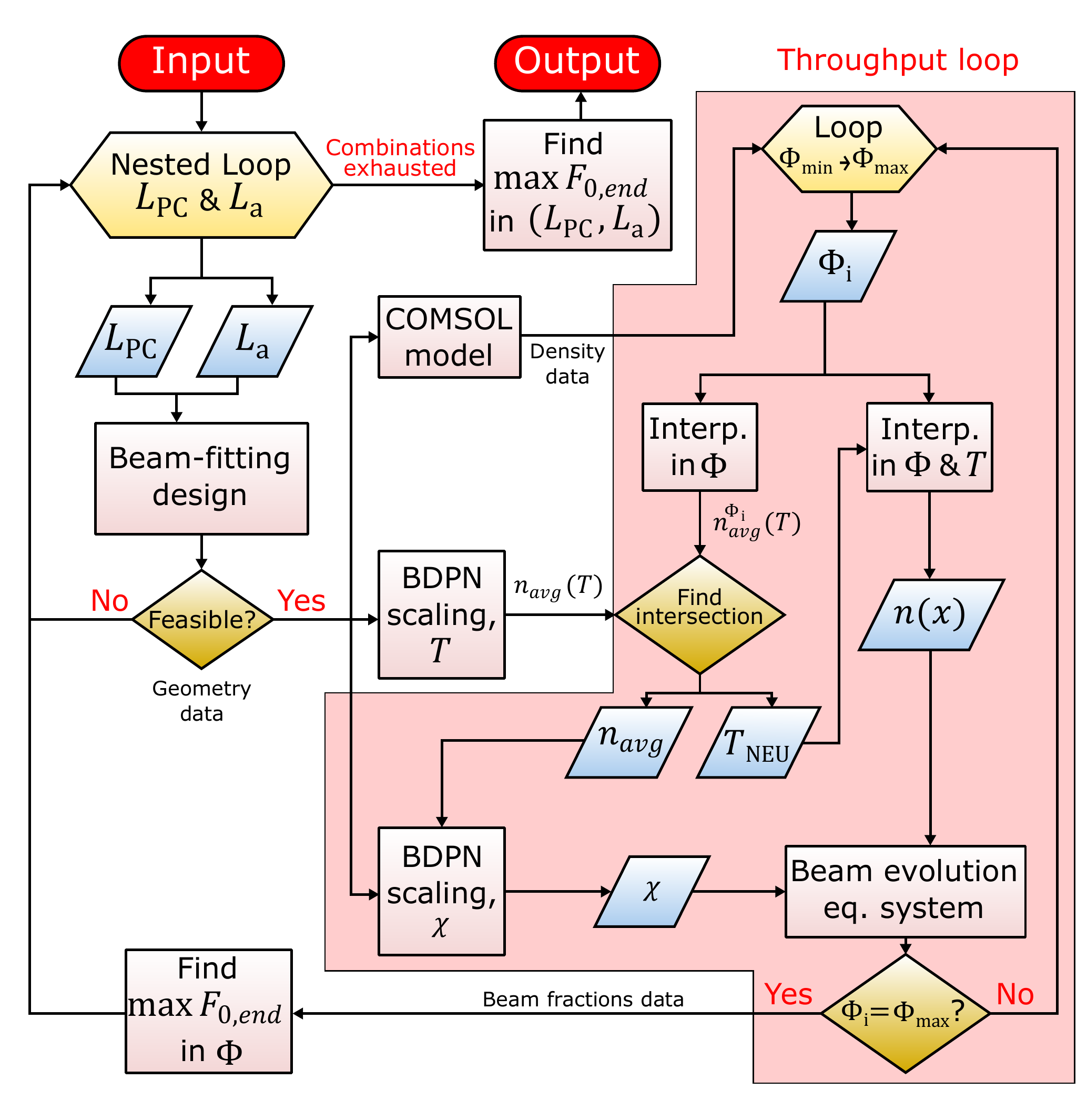}
	\label{fig:algorithm}
\end{figure*}
\end{document}